\documentclass[12pt]{article}

\usepackage{amsmath,amssymb,graphicx}
\usepackage[utf8]{inputenc}
\usepackage[margin=1.0in]{geometry}
\usepackage{csquotes}
\usepackage{caption}
\usepackage{subcaption}
\usepackage{mathrsfs}
\usepackage{xcolor}
\usepackage{lineno}
\usepackage{mathtools}
\usepackage{amsmath}
\usepackage{amssymb} 
\usepackage[colorlinks=true, linkcolor=blue, citecolor=red, urlcolor=blue]{hyperref}
\usepackage{authblk}

\newcommand{\mcL}{\mathcal L}

\newcommand{\ba}{{\boldsymbol a}}
\newcommand{\bbe}{{\boldsymbol b}}

\newcommand{\bff}{{\boldsymbol f}}

\newcommand{\bk}{{\boldsymbol k}}
\newcommand{\bx}{{\boldsymbol x}}

\newcommand{\bz}{{\boldsymbol z}}

\newcommand{\be}{{\boldsymbol e}}

\newcommand{\blambda}{{\boldsymbol \lambda}}
\newcommand{\brho}{{\boldsymbol \rho}}
\newcommand{\bnu}{{\boldsymbol \nu}}

\newcommand{\bxi}{{\boldsymbol \xi}}

\newcommand{\reals}{\mathbb R}

\newcommand{\bbP}{\mathbb P}

\newcommand{\bbT}{\mathbb T}
\newcommand{\E}{\mathbb E}

\title{Coarse-graining and stochastic oscillations in a phenomenological model of cell-size homeostasis}

\author[1]{Ethan Levien}
\author[1]{Jessica Rattray}
\affil[1]{Department of Mathematics, Dartmouth College}
\date{\today}

\begin{document}
\maketitle

\begin{abstract}
Within a continuous-time, stochastic model of single-cell size homeostasis, we study how the structure of feedback from size to growth rates and cell-cycle progression shapes overall size dynamics, both within and across cell cycles. We focus on a model in which the feedback from cell size to these other processes occurs only through the size deviations, defined as the difference between the absolute size and the progression through the cell cycle. 
In a linear regime of this model, the dynamics reduce to a stochastically forced simple harmonic oscillator, yielding closed-form expressions for mother-daughter size correlations. We compare these to the higher order regression coefficients that measure the size memory over many generations. Our analysis reveals how the interplay between cell-cycle timing and intrinsic fluctuations shapes the apparent coarse-grained size control strategy, and in-particular, that coarse-grained correlations may not reflect the mechanistic feedback structure. We compare this model to a more commonly used approach where the coarse-grained dynamics are hard-coded into the model; hence, the first order autoregressive model for sizes is a perfect description of the size dynamics and therefore more accurately reflects the feedback structure. 
\end{abstract}
\section{Introduction}

Many cells, tissues, and populations exhibit exponential growth and removal of biomass (by division or death) at some stage of their existence. These systems are typically found to be in homeostasis, meaning that the long-term variance in biomass is bounded, despite the fact that exponential growth is inherently unstable with respect to small fluctuations in growth and removal rates \cite{alon2023systems,karin2016dynamical,billman2020homeostasis,lin2018homeostasis,salman2025emergent}. Therefore, feedback from size to growth and/or removal rates is needed to maintain a stable biomass variance.  The problem of identifying the regulatory processes through which this feedback is implemented is a central challenge in many areas of physiology.

Stochastic oscillations are a generic feature of systems exhibiting homeostasis. To see this, let $m$ be a biomass where the per unit mass growth rate ($\lambda$) and the removal rate ($\nu$) are both subject to small fluctuations. Further, assume that the net growth, $\rho = \lambda - \nu$, is on average zero. Then $dm/dt = \rho m$, and hence $y = \ln m$ obeys
\begin{equation}\label{eq:intro1}
y(t)  = y(0) + \int_0^t\rho(s)ds. 
\end{equation}
In the absence of feedback between $\rho$ and $m$, $\ln m$ is an additive process, and ${\rm var}(y) \sim O(t)$ \cite{touchette2018introduction}. 
In the simplest case, we have linear dynamics for $\rho$:
\begin{equation}\label{eq:intro2}
\frac{d}{dt}\rho = -a\rho - fy + \xi
\end{equation}
where $\xi$ is a noise term.  
The dynamics of $y$ then map to a \emph{simple harmonic oscillator} (SHO) subject to white noise forcing. The natural frequency and damping ratio are given by $\omega_0 = b$ and $\eta = a/f$, respectively.
It is important to understand the consequences of stochastic oscillations for model identification and coarse-graining in the context of physiology.

This paper concerns models of homeostatic feedback in dividing cells that accumulate biomass exponentially between divisions. The question of how homeostasis is maintained in this context has played a central role in cell biology for decades \cite{willis2020limits,bjorklund2019cell,ho2018modeling}. Because cell size depends on both biomass accumulation (a proxy for anabolic metabolism) and the timing of division, questions about homeostasis necessarily connect two fundamental processes: metabolism and the cell cycle. There is now a vast body of literature focusing on the genetic and molecular components of cell-cycle control. These include the Retinoblastoma (Rb) gene family \cite{knudson1971mutation}, E2F transcription factors, and cyclin-dependent kinases in Eukaryotes,
 and proteins such as FtsZ and DnaA in bacteria \cite{murray1991controls,lutkenhaus1993ftsz}. 
However, there remain many open questions about how the cell cycle and metabolic pathways interact. In particular, a major question is whether homeostatic control of growth occurs through regulation of growth rates (reducing the growth of larger cells), generation times (accelerating division in larger cells), or both \cite{si2019mechanistic,cadart2018size,wang2021regulation,jones2001growth,salman2025emergent}.

\begin{figure}[h!]
\centering
\includegraphics[width=0.9\textwidth]{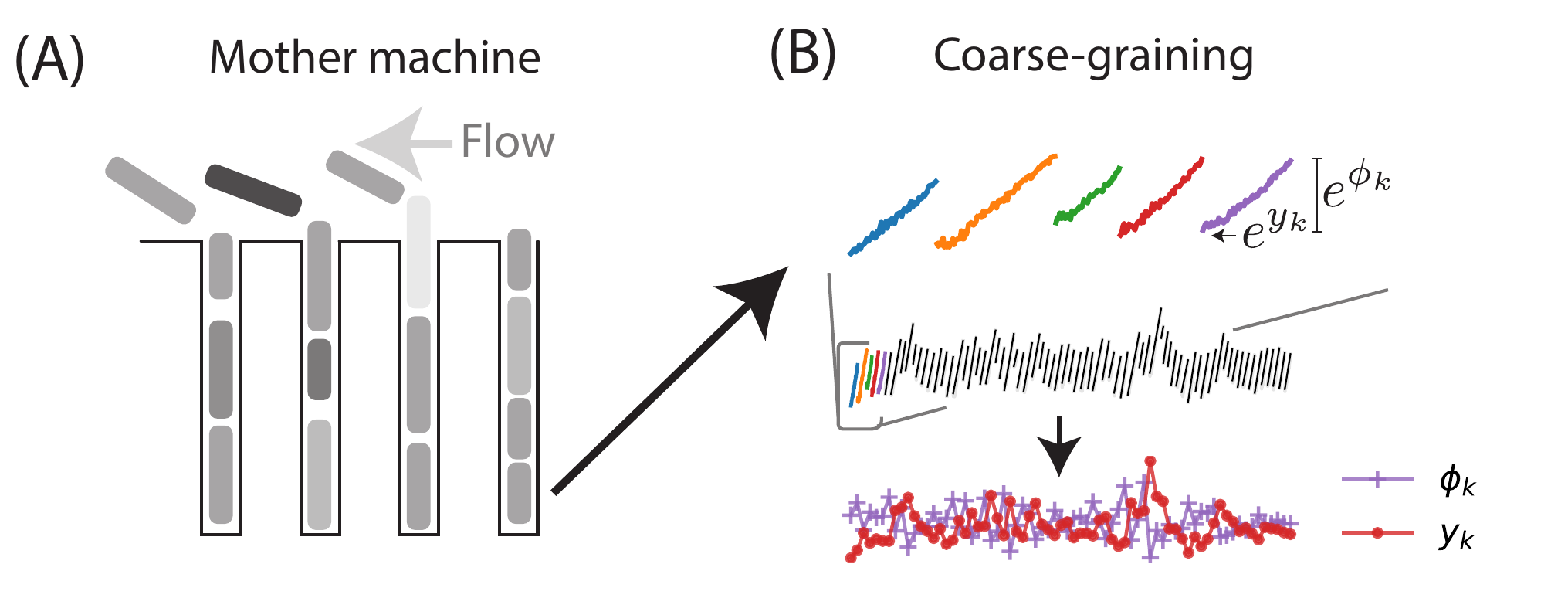}
\caption{
 (A) A diagram of the mother machine, a device used to measure single-cell size. Single bacterial cells (gray rectangles) are trapped in microfluidic channels and recorded for many generations. (B) An example of \emph{E. Coli} data collected from the mother machine \cite{tanouchi2017long} and an illustration of coarse-graining continuous size dynamics. In this case each cell cycle's continuous growth curve is replaced with two variables $\phi_k$ and $y_k$ representing the (log) fold change in size and (log) initial size. }
\label{fig:1}
\end{figure}

We can understand these two processes -- growth and cell-cycle progression -- as corresponding to growth and removal, since at the end of the cell-cycle a cell divides, removing a portion of the biomass. However, a distinctive feature of single cell dynamics, compared to models of other homeostatic systems in physiology, is the inherent asymmetry between growth and removal processes introduced by the discrete nature of division. We therefore cannot necessarily treat these processes on equal footing and expect to disentangle their roles in maintaining homeostasis. Indeed, in many previous models of cell-size homeostasis, one selects either growth or the generation time to be subordinate to the other (usually growth) \cite{levien2025stochasticity,hein2024asymptotic}.  

We find it interesting to explore an alternative class of models that, after a transformation of the state space, have the symmetry properties enjoyed by the SHO-type model introduced above. To this end, we will introduce a model of cell size homeostasis which can be mapped to an SHO by defining a certain linear combination of the (log) size and cell-cycle, called the size-deviations. This model can realize a range of cell-size control strategies and coarse-grained dynamics, making it a natural test case for investigating questions about coarse-graining homeostatic dynamics.

Our model will be connected to previous discrete time models of size dynamics, which average over the continuous dynamics within the cell-cycle to obtain a discrete autoregressive process (see Figure \ref{fig:2} (B)). These models are evaluated based on measurements of single-cell size obtained from the \emph{mother machine}, a microfluidic device where single lineages are imaged in microscopic channels \cite{10.1016/j.cub.2010.04.045,10.7554/elife.88463,10.1101/2023.11.23.568485,10.1371/journal.pone.0236534}. One motivation for working with discrete time models is that measurements within the cell-cycle from these devices are subject to high measurement noise, although unlike buoyant mass measurements from the SMR, these devices are able to collect hundreds of cells per experiment.  A challenge in the field is to link the continuous-time and discrete auto-regressive models \cite{luo2023stochastic,hein2024asymptotic,kohram2021bacterial}.

The organization of this paper is as follows. 
In Section \ref{sec:mod}, after providing some background, we introduce a continuous-time model in which biomass growth and the cell cycle are coupled by means of a hidden variable. The model is defined by the biomass growth rate, the rate at which the cell cycle progresses, and their dependence on a hidden variable that evolves stochastically. In the case of symmetric division, the model can be transformed to a system of SDEs on $\bbT^1 \times \mathbb{R}^{d+1}$.\footnote{$\mathbb{T}^d$ denotes the $d$-torus and hence $\mathbb{T}^1$ is a circle.} 

Section \ref{sec:linear} presents some results on the linear version of this model. Under some restrictions, the model maps to a simple harmonic oscillator with white noise forcing. 
In Section \ref{sec:cg}, we study the coarse-grained dynamics that come from averaging over the cell-cycle. In particular, we obtain an approximate formula for the coarse-grained autoregressive dynamics of cell size in terms of the damping factor and natural frequency. The dependence of the coarse-grained dynamics on the feedback strength differs qualitatively depending on whether we are in the under- or overdamped regimes and may not reflect the mechanistic structure of feedback.
Finally, in Section \ref{sec:noage}, we compare our model to a class of previously studied models where the cell-cycle progress and growth variables cannot be treated symmetrically.

\section{Background and Model}\label{sec:mod}

\subsection{Background: Discrete time models of cell-size homeostasis}\label{sec:back}
This section provides a brief background on previous work on phenomenological modeling of cell-size control.  We refer to \cite{ho2018modeling} for a review of models of cell-size control and a discussion of more mechanistic modeling approaches.

 
The simplest model in which we can identify a clear regime in which cell size is regulated is a first order auto-regressive model (AR(1) process). Due to the exponential nature of cell growth, it is natural to define an AR(1) process for the log-size at birth, denoted $y_k = \ln m_k$  \cite{ho2018modeling,amir2020thinking,amir2014cell}. 
Assuming that a cell accumulates size at a random rate $\lambda_k$ and has an inter-division (generation) time $\tau_k$, we have 
\begin{equation}
m_{k+1} = \frac{1}{2}m_ke^{\bar{\lambda}_k \tau_k}. 
\end{equation}
Taking the log yields  
\begin{equation}\label{eq:lnmar}
y_{k+1} = -\ln(2) +  y_k + \phi_k
\end{equation}
where $\phi_k = \bar{\lambda}_k \tau_k$. If $\phi_k$ is uncorrelated with initial size, the variance of $\ln m_k$ (and hence $m_k$) will diverge. For this to be an AR(1) process, we take
\begin{equation}\label{eq:arphi}
\phi_k = \ln(2) - \alpha y_k + \xi_{\phi}
\end{equation}
where $\xi_{\phi}$ is a Gaussian random variable which is uncorrelated from $\{y_j\}_{j\le k}$. The AR(1) process for $y_k$ is therefore
\begin{equation}\label{eq:yar}
y_{k+1} =(1-\alpha)y_k + \xi_{\phi}. 
\end{equation}
In this first order model, the regression coefficient $\alpha$ of log initial size on the log fold change in size is known as the \emph{cell-size control parameter}  \cite{cadart2018size,amir2014cell,xia2020pde}. From elementary theory of linear regression, 
\begin{equation}\label{eq:alpha_def}
 \alpha = -\frac{{\rm cov}(\phi_k,y_k)}{{\rm var}(y_k)}.
\end{equation}
A special significance is ascribed to the values $\alpha=1/2$ and $\alpha=1$ in the literature because they correspond to the adder case (adding an approximately fixed increment between birth and division) and the sizer case (dividing at a fixed size threshold), respectively \cite{campos2014constant,taheri2015cell,soifer2016single,amir2014cell,witz2019initiation,si2019mechanistic}.

We can see Eq. \ref{eq:lnmar} as a discrete version of Eq. \ref{eq:intro1}, with $\phi_k$ playing the role of $\rho$. The question of whether feedback occurs through the growth rate or generation time is analogous to the question posed in the introduction about whether feedback occurs via the growth or removal rate.  In \cite{cadart2018size}, the authors propose using the scaled regression coefficient of initial (log) size on the growth rate, $\alpha_{\lambda}$, as a metric for the contribution of growth rate modulation to size homeostasis, namely
\begin{equation}\label{eq:alpha_lam}
 \alpha_{\lambda} = -\frac{{\rm cov}(\lambda_k,y_k)}{{\rm var}(y_k)}\E[\tau_k].
\end{equation}
They define the regression coefficient of generation time on initial size, denoted $\alpha_{\tau}$, similarly.  When $\alpha_{\lambda}$ is computed from  \emph{E. Coli} data, it is found that $\alpha_{\lambda} \approx 0$ suggesting feedback is predominately through the cell-cycle progression rate. This has also recently been shown to hold for biomass measurements of lymphocytic leukemia cells (the L1210 cell line)\cite{levien2025stochasticity}.  

\subsection{Size and cell-cycle dynamics}\label{sec:modelgeneral}
We now describe our model, which is a certain stochastic process $(y(t),\theta(t),\bx(t)) \in \reals \times [0,1) \times \reals^d$. Here $y = \ln m$, $\theta \in \bbT^1 \equiv [0,1)$ is a cell-cycle variable and $\bx(t) \in \reals^d$ is an unobservable quantity. $\theta$ should be interpreted as an abstract representation of the position in the cell-cycle, triggering division upon reaching the threshold $\theta=1$.  In principle, $\theta$ can also be connected to specific biological quantities (see Appendix \ref{app:models}), or, more generally, $\theta$ could be obtained by phase reduction of a higher dimensional stochastic limit-cycle  (see \cite{houzelstein2025generalized} for details on phase reduction for stochastic oscillators). We do not pursue this here, but rather take for granted that such a variable is meaningful. 

We then assume there are two rates $\nu,\lambda: \reals^d \to \reals$ such that $y$ and $\theta$ obey
\begin{align}\label{eq:model}
\frac{d}{dt}\theta &= \nu(\bx)\\
\frac{d}{dt}y &= \lambda(\bx)
\end{align} 
To be biologically meaningful, $\nu$ and $\lambda$ should be strictly positive; however, we will consider a small noise regime where negative rates are possible but occur with exponentially small probability.

\begin{figure}[h!]
\centering
\includegraphics[width=0.9\textwidth]{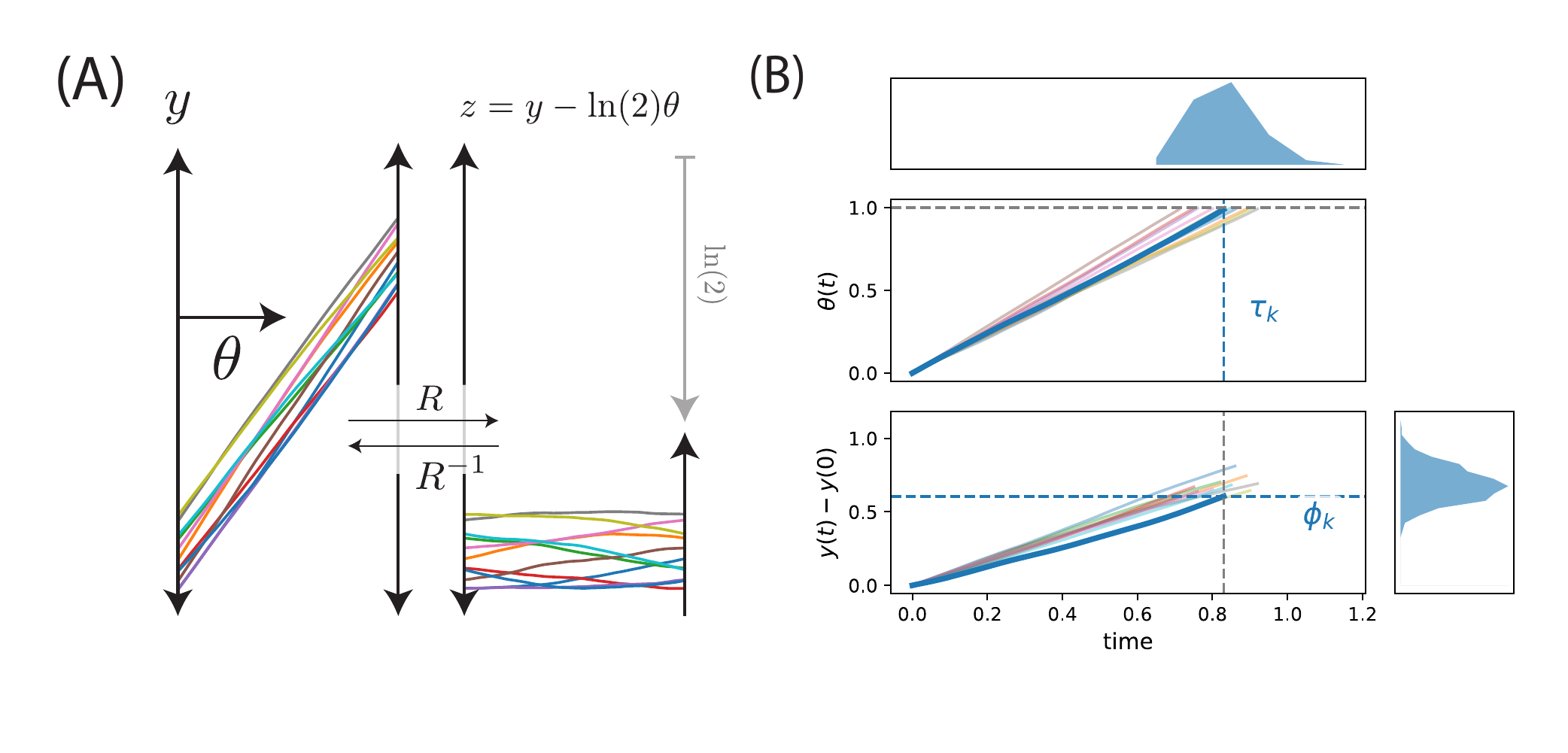}
\caption{
  (A) The transformation from a system with jumps to a continuous system on the cylinder is possible when division is symmetric. As explained in Section \ref{sec:mod}, this can be understood as an invertible transformation $R$ of a certain quotient space on which the original size dynamics live. (B) The dual relationship between the coarse-grained variables $\phi_k$ and $\tau_k$. 
}\label{fig:2}
\end{figure}

We assume perfectly symmetric division; hence, Eq. \ref{eq:model} is supplemented by the division condition, which subtracts $\ln(2)$ from $y$ every time we cross the boundary $\theta=1$.
The assumption of symmetric division allows us to remove the jumps in $y$ and recast Eq. \ref{eq:model} as a continuous system via an invertible transformation $R:  \bbT^1 \times \reals^{d+1} \to \bbT^1 \times \reals^{d+1}$ which maps $y$ to a size-deviation variable $z$ according to
 \begin{equation}
y \mapsto z = y - \ln(2) \theta,
\end{equation}
as illustrated in Figure \ref{fig:2} (A).
 The new variable $z$ obeys 
\begin{equation}
\frac{d}{dt}z = \rho,\quad \rho = \lambda - \ln(2) \nu. 
\end{equation}
Thus, the model has the form of the model described in the introduction 
with $\lambda$ and $\ln(2) \nu$ playing the roles of biomass growth and removal, respectively. We next specify how feedback from $z$ and $\theta$ (or $y$ and $\theta$) to $\bx$ is implemented.

\subsection{Coarse-grained variables}
For a cell born at time $t=0$, the coarse-grained variables $\phi_k, \tau_k, y_k$ and $\lambda_k$ defined in Section \ref{sec:back} are defined in terms of the continuous variable by the relations 
\begin{align}
\phi_k &= \int_0^{\tau_k}\lambda(\bx(s))ds\label{eq:phikcg}\\
1 &=\theta(\tau_k) =  \int_0^{\tau_k}\nu(\bx(s))ds\\
y_{k+1} &= y_k + \phi_k = z(0) + \int_0^{\tau_k}z'(s)ds. 
\end{align}
These quantities are shown in Figure \ref{fig:2} (B).
The fact that our model specifies the dynamics of $\nu$ and $\lambda$ means that $\tau_k$ and $\phi_k$ are difficult to compute, as these involve solving a first passage time problem.  On the other hand, formulating the model in this way makes it easier to study the dynamics of $z$ and illustrates some of the nuances which come with coarse-graining a continuous model. In Section \ref{sec:noage}, we will look at a model where the distribution of $\phi_k$ is ``hard-coded'' into the dynamics, making the coarse-graining more straightforward but leading to more complex dynamics for $z$. Part of our motivation was to lay out the distinction between these approaches.

\subsection{SDE model for hidden variable}
It is possible to implement a wide range of models if we allow ourselves enough flexibility in how $\bx$ is modeled (see Section \ref{sec:noage}). We mostly focus on the case where $\bx$ fluctuates continuously across cell-divisions. In such a model the coarse-grained dynamics are an emergent property and can depend on the model parameters in potentially complex ways. This is a  natural context to pursue questions about the identifiably of biological model parameters from coarse-grained data, even if the parameters in our model are still phenomenological. 
To this end, we consider the case where the $\bx$ dynamics are given by 
\begin{equation}\label{eq:xsde}
\frac{d}{dt}\bx = -A\bx - \bff(\theta,z) + \Gamma  \bxi
\end{equation}
where $A \in \reals^{d\times d}$ is a diagonalizable matrix with all positive eigenvalues, $\Gamma$ is $d \times d$ a non-singular matrix and $\bxi$ is a $d$-vector of white noises.

For fixed $\theta$ and $z$, $\bx$ will fluctuate around a $(\theta,z)$-dependent fixed point $\bbe(\theta,z) = -A^{-1}\bff(\theta,z)$ and it is assumed that Eq. \ref{eq:xsde} defines a stationary, ergodic It\^{o} diffusion process. 
The joint probability density $p_t(\theta,z,\bx)$ of the full process $(\theta(t),z(t),\bx(t))_{t\ge 0}$ obeys a Fokker-Planck (FP) equation 
\begin{equation}\label{eq:fp}
\frac{\partial}{\partial t}p_t = \mcL(\theta,z)p_t -   \nu \frac{\partial}{\partial \theta}p_t  - 
\rho  \frac{\partial}{\partial z}p_t 
\end{equation}
where $\mcL(\theta,z)$ acts on the $\bx$-coordinate of a suitable function $p$ according to
\begin{equation}
(\mcL(\theta,z)p)(\theta,z,\bx) = \nabla \cdot \left[ (A (\bx - \bbe(\theta,z))p(\theta,z)\right] +  \frac{1}{2}{\rm Tr}\left(\Gamma \Gamma^{T}  \nabla^2p(\theta,z,\bx)\right).
\end{equation}
The equation is supplemented by the periodic boundary conditions, $p_t(0,z,\bx) = p_t(1,z,\bx)$. 
We note that the lack of dependence of $\bx$ on the initial size is crucially important; this is the assumption which will allow us to study the dynamics as a continuous dynamical system.\footnote{For a general $f$, this is not as restrictive as it may seem; since $f$ depends on $\theta$,  with a suitable chosen nonlinearity we can force $\bx$ to evolve quickly at the beginning of the cell-cycle in such a way that the dependence is primarily through the initial size. Despite this, we will focus on versions of the model where this is not possible, leaving such generalizations to future work.}

A special case of Eq. \ref{eq:fp} is when $\mcL$ is independent of $\theta$ (equivalently, $\bff$ is independent of $\theta$). Note that this \emph{does not} mean the $\bx$ dynamics are unaffected by $\theta$, rather it implies that all the effects of $\theta$ occur through the deviations of $y$ from the intrinsic cell-cycle variable, and as a result, the dynamics are smooth across cell-divisions when $\bff(z,\bx)$ is smooth. In this case, the density marginalized over $\theta$, $\bar{p}_t(z,\bx) = \int_0^1p_t(\theta,z,\bx)d\theta$ obeys the averaged FP equation 
\begin{equation}\label{eq:fpavg}
\frac{\partial}{\partial t}\bar{p}_t + \rho  \frac{\partial}{\partial z}\bar{p}_t  =  \mcL(z)\bar{p}_t. 
\end{equation}
To obtain this equation, we have used $\int_0^1\nu p_t(\theta,z,\bx)d\theta =0$ due to the boundary conditions. 

\section{Linear regime and stochastic oscillations}\label{sec:linear}
Here we study a special case of the model which maps to the SHO. When $\nu$, $\lambda$ and $\bff$ are all linear, our model becomes
\begin{equation}\label{eq:linear}
\frac{d}{dt}
\begin{bmatrix}
\theta \\ y \\\bx\end{bmatrix} =
\begin{bmatrix}
\nu_0 \\
\lambda_0 \\
0\end{bmatrix}+
\begin{bmatrix}
0 & 0 & \bnu_x^T \\
0 & 0 & \blambda_x^T \\
-\bff_\theta & -\bff_y & -A
\end{bmatrix}
\begin{bmatrix}
\theta \\
y \\ \bx\end{bmatrix}+\begin{bmatrix}
{\bf 0}^T \\{\bf 0}^T \\\Gamma\end{bmatrix}\xi
\end{equation}
which in the $z$-coordinates is 
\begin{equation}\label{eq:1dz}
\frac{d}{dt}
\begin{bmatrix}
\theta \\ z \\\bx\end{bmatrix} =
\begin{bmatrix}
\nu_0 \\
\rho_0 \\
0\end{bmatrix}+
\begin{bmatrix}
0 & 0 & \bnu_x^T \\
0 & 0 & \brho_x^T\\
-\tilde{\bff}_\theta & -\bff_y & -A
\end{bmatrix}
\begin{bmatrix}
\theta \\
z \\ \bx\end{bmatrix}+\begin{bmatrix}
{\bf 0}^T \\{\bf 0}^T \\\Gamma\end{bmatrix}\xi
\end{equation}
The transformed coefficients are related to the original coefficients by
\begin{align}
\brho_x &= \blambda_x - \ln(2) \bnu_{x}\label{eq:uzdef}\\
\rho_0 &= \lambda_0 - \nu_0 \ln(2)\label{eq:delta}\\
\tilde{\bff}_{\theta} &= \bff_{\theta} + \ln(2)\bff_{y}\label{eq:btildef}.
\end{align}
Remember that the equation for $\theta$ is unaffected by the coordinate transformation $R$. 
Note that in general, the dynamics are still non-linear due to the non-Euclidean state space, but we will derive an analytical formula in a linear limit where $\nu$ is constant. 

We now make two simplifying assumptions under which the dynamics reduce to an SHO. We assume $\tilde{\bff}_{\theta}=0$ (corresponding to the case where $\bx$ is independent of $\theta$ in Eq. \ref{eq:xsde}) the dynamics of $z$ decouple from the cell cycle as explained in the previous section. Second, we set $A = aI$ for a scalar $a$ (the relaxation of the hidden variable is isotropic). Note that without loss of generality we may set $\rho_0=0$ since this can be achieved by translating $(\bx,z)$. In Appendix \ref{app:evalgeneral} we discuss the eigenvalue problem in the general case.

Taking the second derivative of $z$ yields 
\begin{align}
\frac{d^2}{dt^2}z &=  \brho_x^T\frac{d}{dt}\bx = -a   \brho_x^T\bx  -   \brho_x^T\bff_{y} z  +   \brho_x^T\Gamma \bxi.
\end{align}
Since $\xi' =   \brho_x^T \Gamma \bxi$ is a new white noise term with magnitude $\sigma = || \Gamma\brho_x||_2$ and $\brho_x^T\bx = dz/dt$, we can rewrite this as
\begin{equation}\label{eq:sho}
\left[\frac{d^2}{dt^2} + 2 \eta \omega_0 \frac{d}{dt} + \omega_0^2\right]z  = \sigma \xi'
\end{equation} 
where the frequency $\omega_0$ and damping ratio $\eta$ are 
\begin{align}
\omega_0^2 &=  \bff_y^T\brho_x\label{eq:omega0}\\
 \eta &= \frac{a}{2\sqrt{\bff_y^T \brho_x}}\label{eq:eta}. 
\end{align}
 In order for the origin to be a stable fixed point, we require $a>0$ and $\bff_y^T \brho_x > 0$. 
The condition $\bff_y^T  \brho_x > 0$ can be rewritten as 
\begin{equation}\label{eq:stability_cond}
\bff_y^T\blambda_x > \ln(2) \bff_y^T \bnu_{x}
\end{equation}
If $\bnu_{x} =0$ this becomes $\bff_y^T\blambda_y > 0$ which intuitively makes sense: the feedback needs to compensate for the fluctuations. 
However, if $\bnu_{x}> 0$, the influence $\bx$ has on the cell-cycle progression rate can also account for some of the correlations.

From the perspective of the $z$-dynamics (but not the coarse-grained dynamics; see below) the hidden variable $\bx$ can be replaced with a one-dimensional hidden variable $\tilde{x} = \brho_x^T\bx$ and the joint dynamics of $(z,\tilde{x})$ are well understood. In particular, the solution of the averaged FP equation (Eq. \ref{eq:fpavg}) for $(z,\tilde{x})$ is known as the \emph{Smoluchowski} or \emph{Kramers} equation, which can be represented by weighted Hermite polynomials \cite{risken1989fokker}. Moreover, the stationary density $\bar{\pi}(z,\tilde{x}) = \lim_{t \to \infty}\bar{p}_t(z,\tilde{x})$ is a Multivariate Gaussian with a diagonal covariance matrix
\begin{equation}\label{eq:shoSigma}
\Sigma = \begin{bmatrix}
 \sigma^2a^{-1} & 0 \\
 0 & \sigma^2\omega_0^{-2}a^{-1}
\end{bmatrix}. 
\end{equation}
This implies that size deviations and growth rates are independent in the stationary distribution.

 The phenomena of decoupling between cell-size and growth rates in models of single-cell size homeostasis was discovered in  ref \cite{hein2024asymptotic}, where the authors show that growth rate and cell-size decouple asymptotically along a lineage and in a snapshot of a growing population. More general results on decoupling between size and growth rates are given in \cite{levien2025size}. Here, the decoupling is instead between the size deviations from the growth rate. The two notions of decoupling are equivalent to the decoupling of size and growth rates when $x$ is one-dimensional, and hence the size growth rate and cell-cycle progression rate are perfectly correlated. However, in Section \ref{sec:cg} we will see that the notions of decoupling considered in \cite{hein2024asymptotic,levien2025size} do not in general hold even when the there is no feedback from size to the growth rates.

   \begin{figure}[h!]
\centering
\includegraphics[width=1.\textwidth]{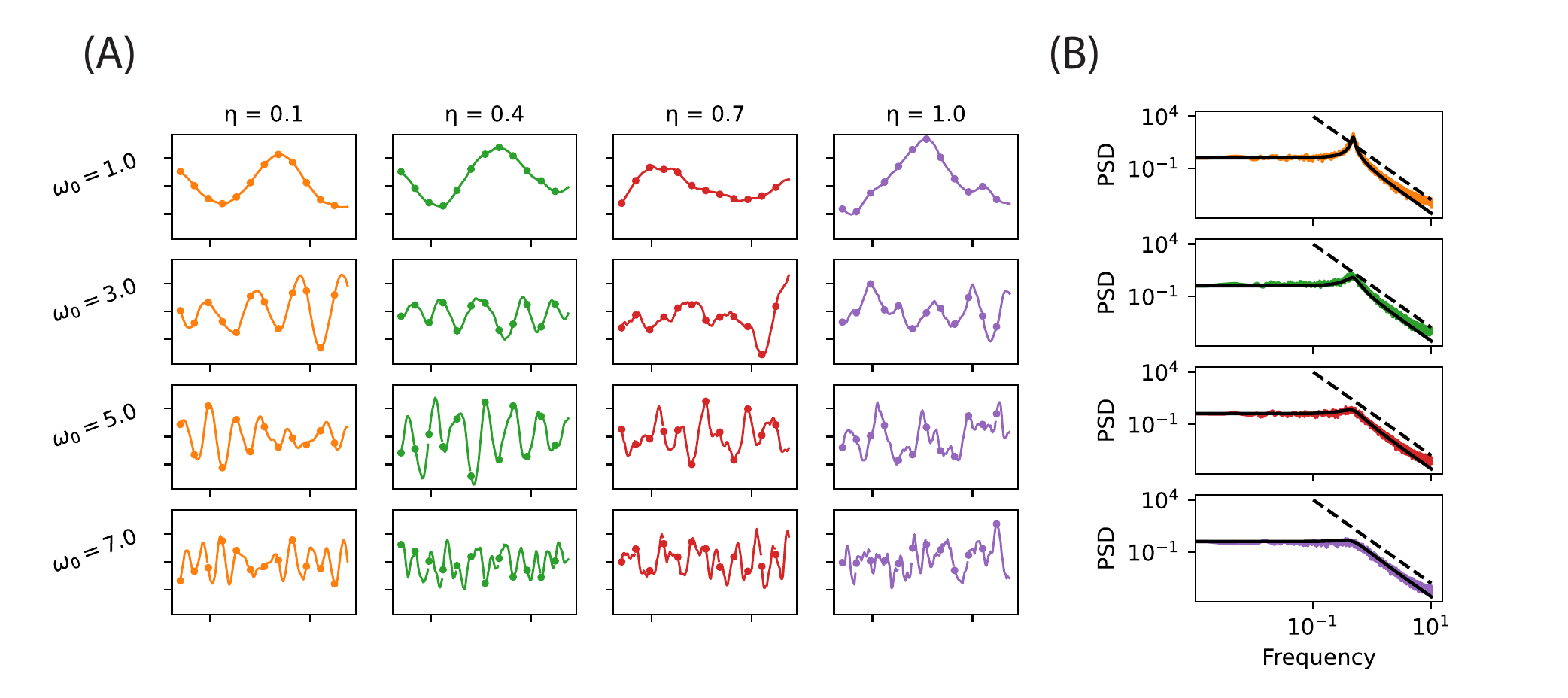}
\caption{
        (A) Representative $z$ dynamics for varying $\omega_0$ (top to bottom) and $\eta$ (left to right).  The dots show where the cell divisions occur.
        (B) Corresponding power spectral densities (PSDs) computed from these trajectories (with $\omega_0=3$). The colors correspond to the trajectories in (A). Details about the parameter values are given in Figure \ref{fig:4}.}
    \label{fig:3}
\end{figure}

An analytical formula for the marginal, stationary autocovariance of $z(t)$, which we denote by $\kappa_z(t)$, is well-known (see \cite{foreman2017fast,anderson1990modeling}) and is given by \begin{equation}\label{eq:acf1d}
\kappa_z(t) \equiv {\rm cov}(z(0),z(t)) = \frac{\rho_x^2\sigma^2}{\omega_0^{2} a}e^{-t \eta \omega_0/2} C(t)
\end{equation}
where $ \varrho  = \sqrt{|1-\eta^2|}$ and 
\begin{equation}
C(t) = 
\begin{cases}
\cosh( \varrho \omega_0t) + \frac{\eta}{ \varrho } \sinh(\varrho \omega_0t), &  \eta >1 \\
2(1 + \omega_0 t), & \eta = 1 \\
\cos(\varrho \omega_0t) +  \frac{\eta}{ \varrho }  \sin(\varrho \omega_0t), &  \eta < 1
\end{cases}
\end{equation}
Taking the Fourier transform gives the power frequency spectrum \cite{foreman2017fast}, which is given by  
\begin{equation}\label{eq:Somega}
S(\omega) = \sqrt{\frac{2}{\pi}} \frac{ S_0 \omega_0^4}{\left(\omega^2 - \omega_0^2\right)^2 + \omega^2 \omega_0^2\eta^2},
\end{equation}
where $S_0$ is independent of $\omega_0$ and $\eta$. The critical case, $\eta=1$, corresponds to the so-called \emph{Mat\'{e}rn} covariance function, which was used to model growth rate fluctuations in \cite{levien2025stochasticity}. Hence, this provides some motivation for using such a kernel in the analysis of single-cell biomass data.  Eq. \ref{eq:Somega} is compared to simulations in Figure \ref{fig:3} (A) in the case $d=1$. In this figure we can also see the qualitative differences between regimes of high and low $\eta$ and $\omega_0$.

\section{Connecting continuous to coarse-grained dynamics}\label{sec:cg}

\subsection{Coarse regression coefficients for the SHO limit}
It is difficult to obtain analytical results for coarse-grained dynamics from the continuous-time model because it involves solving a first passage time problem for a Gaussian process, for which only approximations are available \cite{reith2023approximate}.  However, we can derive analytical results when $\bnu_{x}=0$.  For simplicity, we begin with $d=1$. In this case, the division events occur at fixed intervals $\nu_0$ and their position in time has no influence on the $y$ dynamics; thus, the coarse-grained process $(z_k,\lambda_k)$ obtained by sampling the log size and growth rate at division is Markovian (and is an AR1 process). Unfortunately, this is not the experimentally observed coarse grained process, since we observe the cell-cycle averaged growth rate $\bar{\lambda}_k$. In the limit $\nu_x \to 0$, $\bar{\lambda}_k = \phi_k\nu_0$ and $\phi_k$ is determined deterministically by $z_{k}$ and $z_{k-1}$ according to Eq. \ref{eq:phikcg}.  These observations motivate us to start by examining the marginal process for $z_k$ and comparing this to Eq. \ref{eq:yar}, which is not the correct generative equation for the $z_k$ dynamics due to the inertia coming from $x$.

\begin{figure}[h!]
\centering
\includegraphics[width=1.\textwidth]{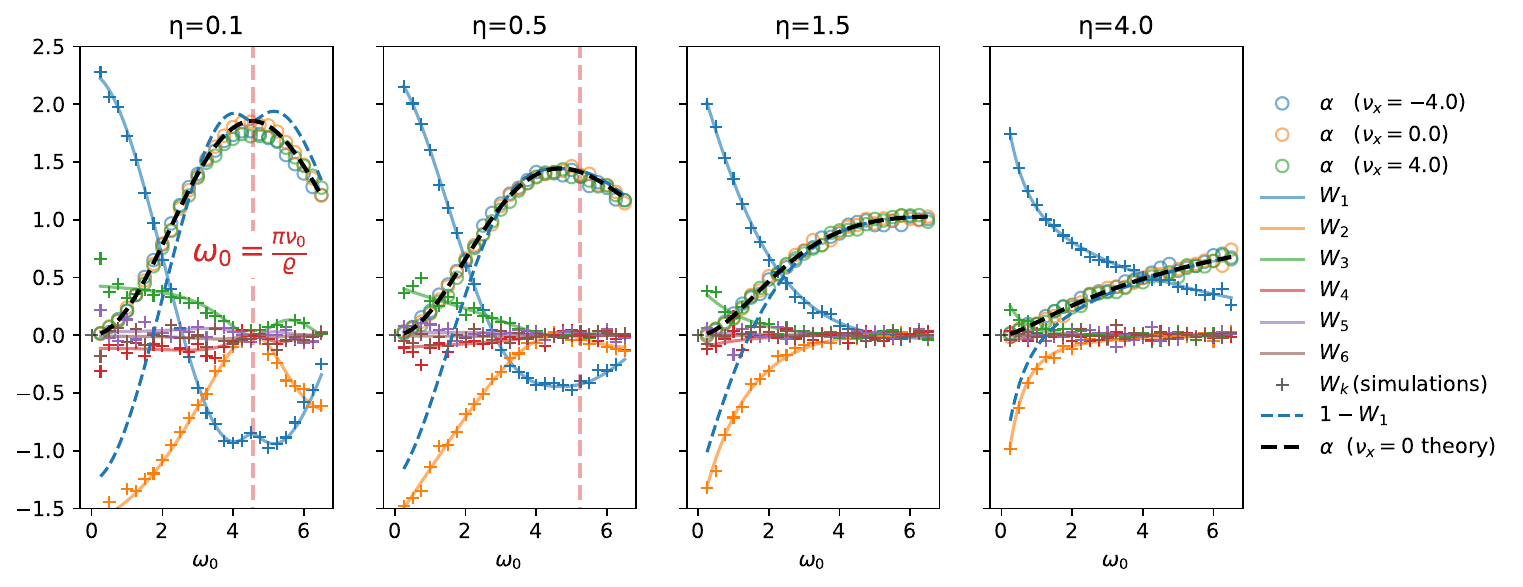}
\caption{ The coarse-grained regression coefficients, $\alpha$ and $W_{i}^{(l)}$, as a function of $\omega_0$ for different $\eta$. The theoretical values are for $\nu_x=0$ and the circles show simulated data. We plot $\alpha$ because it is equal to $1-W_{1}^{(1)}$. and hence compared to $W_1^{(l)}$ to see the effects of multigenerational correlations. Here we have used $l=30$, although only the first $\approx 5$ coefficients are non-negligible. Other parameters are $\rho_0=0$ and $\sigma$ is determined by $k_z(0) = 10^{-3}$ via Eq. \ref{eq:acf1d}. } \label{fig:4}
\end{figure}

Let $t=0$ be the beginning of the $k$th cell-cycle. Then, using $z(1/\nu_0)= z(0) + \phi_k + \ln (2)$ and $z(0) = y_k$ we have   \begin{align}
  {\rm cov}(y_k,\phi_k) &=  {\rm cov}(z(0),z(1/\nu_0) - z(0) - \ln(2))\\
  & = \kappa_z(1/\nu_0) - \kappa_z(0). 
  \end{align}
Thus Eq. \ref{eq:alpha_def} can be rewritten as
 \begin{equation}
 \alpha= 1 - \frac{\kappa_z(1/\nu_0)}{\kappa_z(0)}.
 \end{equation}
In particular, in the underdamped regime
 \begin{equation}\label{eq:alpha_d1}
\alpha = 1 - e^{-\eta \omega_0/(2 \nu_0)} \left(
\cos\left(\frac{\omega_0 \varrho}{\nu_0}  \right)
+ \frac{\eta}{\varrho} \sin\left(\frac{\omega_0 \varrho}{\nu_0} \right)
\right). 
\end{equation}
Eq. \ref{eq:alpha_d1} is compared to our simulations in Figure \ref{fig:4} showing perfect agreement when $\nu_x=0$. Moreover, the formula appears to be a good approximation for $\nu_{x} \ne0$. 

\begin{figure}[h!]
\centering
\includegraphics[width=1.\textwidth]{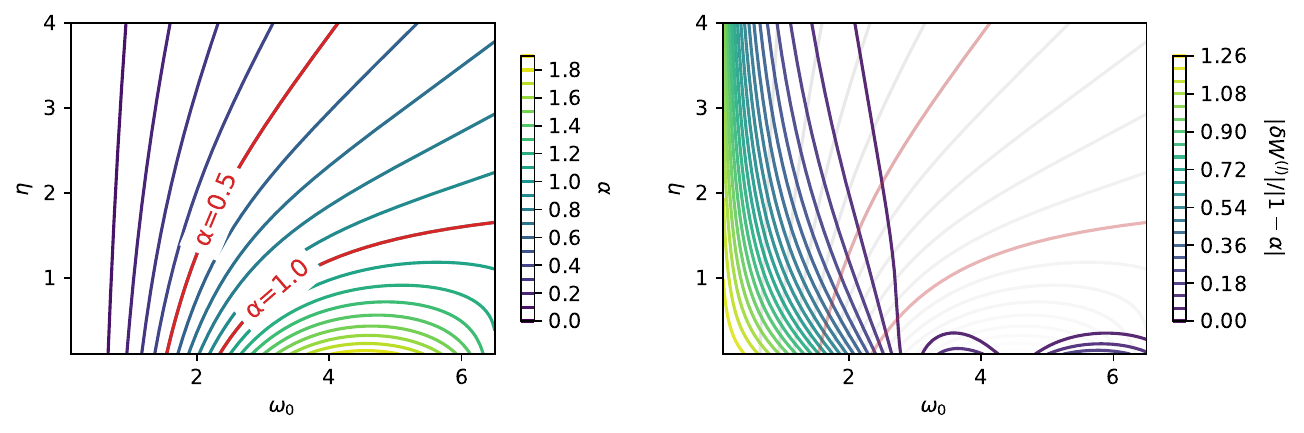}
\caption{ (left) A contour plot of $\alpha$ in the limit $\nu_x=0$ (Eq. \ref{eq:alpha_d1}). The sizer and adder contours are highlighted.  (right) A contour plot of $|\delta W^{(l)}|/|W_1^{(1)}|$ with the $\alpha$ contours in the background.  } \label{fig:5}
\end{figure}

In Figure \ref{fig:5}  we have shown a contour plot of $\alpha(\omega_0,\eta)$ and highlighted the contour for $\alpha=1/2$, which corresponds to the adder strategy observed in many experimental systems.  In the under-damped regime, $\alpha$ is non-monotonic in the feedback strength $f_y$. This is due to the fact that cell-size regulation is maximized when $\omega_0 \varrho/\nu_0 \approx \pi  + k\pi/2$ with $k=1,2,\dots$ as at these values, the cell divisions sample the SHO where the correlations are maximized. This phenomena is well known in the Gaussian processes literature \cite{williams2006gaussian}.

We now look at the higher order regression coefficients.  To express these in terms of the covariance function $\kappa_z(t)$ we use the notation
\begin{equation}
\bz_{k}^{(l)} = \begin{bmatrix}
z_{k-1}\\ 
\vdots \\
z_{k-l}
\end{bmatrix},
\quad \bk_l 
= \begin{bmatrix} \kappa_z(1/\nu_0)\\ \vdots \\  \kappa_z(l/\nu_0)
\end{bmatrix},\quad 
K_l = \begin{bmatrix}  \kappa_z(0) & \cdots &  \kappa_z((l-1)/\nu_0)\\
\vdots & & \vdots\\
\kappa_z((l-1)/\nu_0) & \cdots &  \kappa_z(0)
 \end{bmatrix}.
\end{equation} 
 Using standard formulas for Gaussian process \cite{williams2006gaussian}, we have
\begin{equation}
z_{k}|\bz_k^{(l)} \sim {\rm Normal}( \bk_l ^TK_l^{-1}\bz_k^{(l)},\kappa_z(0) -  \bk_l ^TK_l^{-1}\bk_l).
\end{equation}
Hence for any $l>1$ the $l$th order regression model on the current log size in terms of the previous log sizes is
\begin{equation}\label{eq:zkl}
z_{k} = \sum_{j=1}^{l}W_{j}^{(l)}z_{k-j} + \epsilon_k^{(l)}. 
\end{equation}
with $W_j^{(l)}$ denoting entries of the row vector $\bk_l ^TK_l^{-1}$.  As with Eq. \ref{eq:yar}, it is important to remember that this is a representation of the conditional distributions, not a generative model for the $z_k$ dynamics in the sense that iterating  Eq. \ref{eq:zkl} does not yield sample paths for $z_k$.  Note that when $l=1$, $K_l = \kappa_z(0)$ and $\bk_l = \kappa_z(l/\nu_0)$; therefore, $W_1^{(1)} = 1-\alpha$. 

In Figure \ref{fig:4}, we have compared $1-\alpha$ to $W_1^{(l)}$ with $l = 30$. When $\omega_0 \varrho/\nu_0 \approx \pi$ all but the first regression coefficient vanishes, but for $\omega_0$ greater than or less than the value that satisfies this, the higher order coefficient becomes significant. For sufficiently large $\eta$, the higher order terms vanish since the multigenerational memory of size is lost in this limit.

The regression coefficients in the first order model are related to the regression coefficients in the $l$th order model of $z_k$ according to 
\begin{equation}
W_1^{(1)} = \delta W^{(l)}  + W_1^{(l)}
\end{equation}
where 
\begin{equation}
\delta W^{(l)} =  \sum_{j=2}^l W_j^{(l)}\kappa_z((l-1)/\nu_0)/\kappa_z(0)
\end{equation}
is the standard correction for missing predictors in a linear regression model. The ratio 
\begin{equation}
 \frac{|\delta W^{(l)}|}{|W_1^{(1)}|} = \frac{|\delta W^{(l)}|}{|1-\alpha|}
\end{equation} is therefore a measure of how much of the size correlations are lost by using a first order model. We have plotted these contours in Fig. \ref{fig:5} to be compared with  the $\alpha$ contours. This again illustrates that the multigenerational correlations are only important for low frequencies unless there is very little damping. Importantly, the contours of the fixed cell-size control strategy (adder and sizer are highlighted) pass through the regimes where the first order model is an accurate description, meaning that within this model, we could observe, for example, an adder while the overall variance in sizes is controlled by feedback over multiple generations.

\subsection{Interpolating between control via growth rate and generation time control: the $q$-SHO model} \label{sec:timevgrowth}
We now formulate a version of our model that can interpolate between the limits where homeostatic regularization is achieved via adjustments to growth rate vs. cell-cycle progression.  We call this the $q$-SHO model because it is parameterized by a number $q \in [0,1]$, in addition to the usual frequency and damping ratio.

The model comes from taking $\blambda_x = \be_1$, $\bnu_x = \be_2/\ln(2)$ and
\begin{equation}
\bff_y  = \omega_0^2  \begin{bmatrix} 
q\\
q-1
\end{bmatrix}.
\end{equation}
It follows that $\brho_x = \be_1 - \be_2$ and  $\omega_0^2 = \bff_y^T\brho_x$ for all $q$. We fix $\tilde{\bff}_{\theta}=0$ as we did above, and $\bff_x$ is determined by Eq. \ref{eq:btildef}. Note that the overall scale of $\bx$ can be tuned via $a$ (recall $A=aI$) and $\sigma$. We fix the variance in $z$ and therefore the free parameters are $\omega_0,\eta,q$ and $\sigma$.  

\begin{figure}[h!]
\centering
\includegraphics[width=0.8\textwidth]{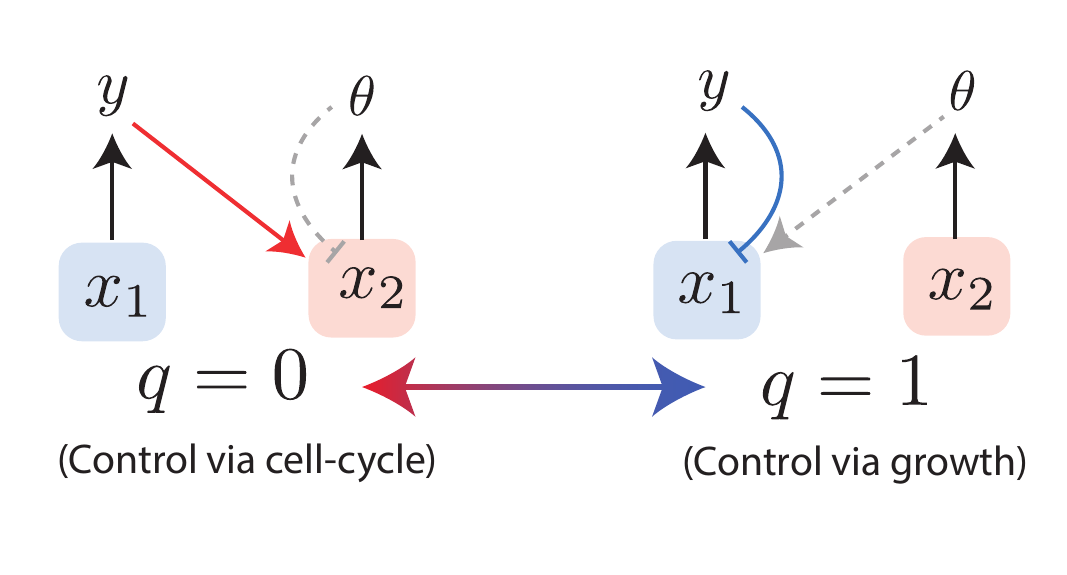}
\caption{A diagram of the feedback structure in the $q$-SHO model. The black arrows represent $\lambda$ and $\nu$, while the colored arrows represent $\bff_y$ and the dashed gray arrows represent $\bff_{\theta}$.
}\label{fig:6}
\end{figure}

A diagram of the model is shown in Figure \ref{fig:6}, which illustrates the structure of feedback in the two limits $q=0$ and $q=1$.  When $q=0$, $\bff_y$ points in the positive $x_2$ direction; hence, large cells have a faster rate of cell-cycle progression. When $q=1$, $\bff_y$ points in the negative $x_1$ direction, and larger cells have smaller growth rates.   The feedback from the cell cycle variable (the dashed line in Figure \ref{fig:6}) is not controlled independently but compensates for the direction of feedback from size to ensure that $\bx$ is only coupled to size deviations.

  \begin{figure}[b!]
\centering
\includegraphics[width=1.\textwidth]{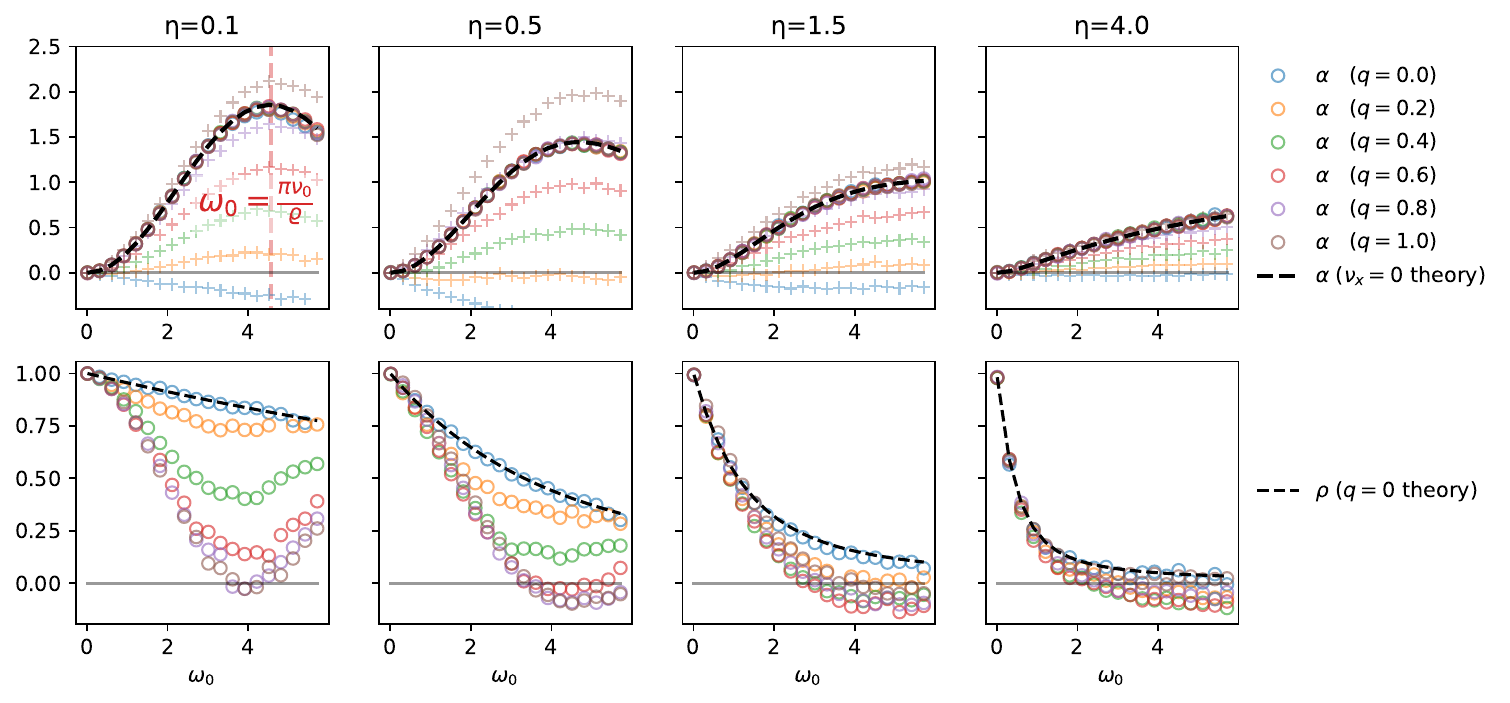}
\caption{
(top) The analytical formula for the cell-size control parameter compared to simulations with different $q$ as a function of $\omega_0$. In the background we have plotted $\alpha_{\lambda}$ (with $+$) to show that values used in Figure \ref{fig:8}. (bottom) $\rho_{\lambda}^{(1)}$, and the formula  given by Eq. \ref{eq:sigmalmlm}. The other parameters are the same as in Figure \ref{fig:4}.
}\label{fig:7}
\end{figure}

Within this model, we found that the relationship between $\alpha$ and the effective frequency and damping factors is the same as for the SHO; see Figure \ref{fig:7}.  When $q=0$, the growth phenotype $x_1$ evolves independently of size as an OU process. In this limit, the marginal growth dynamics are an OU process, although they are not independent of the cell-cycle --see below. The coarse-grained correlations of the growth dynamics in this setting can be approximated by neglecting the high order effects of generation time variability, as is done in \cite{hein2024asymptotic}, and are given by 
\begin{equation}\label{eq:sigmalmlm}
\rho_\lambda^{(j)} = {\rm cor}(\bar{\lambda}_{k+j},\bar{\lambda}_k)
\approx \frac{1}{2} \frac{\left(1 - e^{-\gamma} \right)^2}{
\gamma - \left(1 - e^{-\gamma}\right)}
e^{-(j-1)\gamma},
\quad \gamma = \eta \omega_0 \ln (2)/\E[\lambda]
 \end{equation}
 As shown in Figure \ref{fig:7} this provides an upper bound on the growth rate correlation in our model. This is because size regulation leads to anti-correlations that cancel the heritability intrinsic to the growth process (in our model, these two effects are additive).

\begin{figure}[h!]
\centering
\includegraphics[width=1.\textwidth]{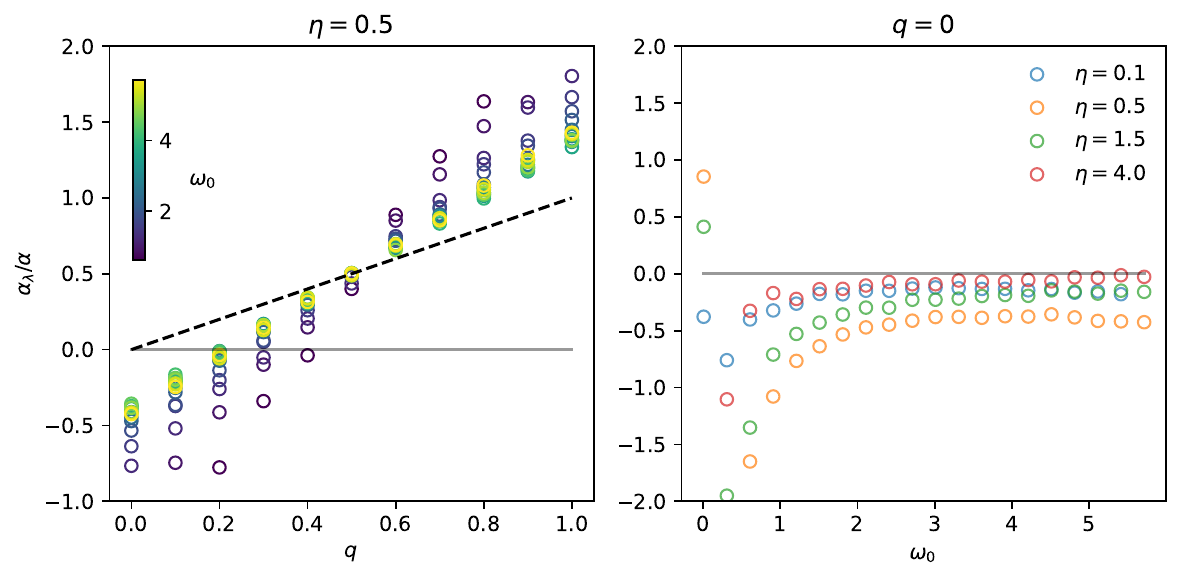}
\caption{  (left) The ratio $\alpha_{\lambda}/\alpha$ for a range of $\omega_0$ values.  (right) For different $\eta$ and $q=0$, the $\alpha_{\lambda}/\alpha$ values as a function of $\omega_0$. The other parameter values are the same as Figure \ref{fig:7}. }\label{fig:8}
\end{figure}

Since the motivation for the $q$-SHO model was to explore the question of whether the effect of growth and generation on size regulation can be deciphered from coarse-grained data, we compared the size regulation parameter $\alpha$ and the measure $\alpha_{\lambda}$ of growth regulation strength. When $\alpha_{\lambda}/\alpha$ is plotted as a function of $q$, it increases monotonically; see Figure \ref{fig:8}. This makes sense because $q$ controls the mechanistic source of feedback, which is well-defined;  however, to interpret $\alpha_{\lambda}=0$ as indicating no role of growth rate in size regulation would be incorrect, as this is non-zero for any finite $\eta$.  This is highlighted in the left panel, where it can be seen that the ratio diverges from $0$ at small frequencies. The reason for this is that the feedback from $y$ to $x_2$ introduces correlations between $x_1$ and $x_2$. In the next section, we compare this to a model where the coarse-grained growth and size correlations vanish, which further elucidates the growth size dependence at $q=0$.

\section{Fluctuating ratio model}\label{sec:noage}

\subsection{General formulation}
As we illustrated in the previous section, the decoupling of growth and cell-cycle progression does not lead to vanishing growth-size correlations. In order for the growth rate to statistically (not just casually) decouple from the cell-cycle, it must be that the cell-cycle progress depends on the growth rate in a manner that cancels with the feedback. This is the case for many previous papers in the literature, wherein cells have no intrinsic notion of age other than their size \cite{hein2024asymptotic,levien2025stochasticity,levien2025size}. 

Such models can be implemented in our notation by replacing the periodic boundary conditions in the model studied above with
\begin{equation}
p_t(0,z,\bx) = \int p_t(0,z,\bx')k(\bx|\bx')d\bx. 
\end{equation}
Here $k(\cdot|\bx')$ is a probability density. 
The models in refs \cite{hein2024asymptotic,levien2025stochasticity,levien2025size} can be viewed as special cases of this class of models, where one of the hidden dimensions, say $x_d$, 
is equal to the ratio $x_d(t) = \lambda/\nu$, and the generator $\mcL$ does not act on this component (thus $x_d$ is constant over the cell-cycle). Then, for a cell born at time $t$ and $t<\tau_k$, $x_d(t) = \phi_k$. To see this, write
\begin{equation}\label{eq:frtheta}
\theta(t)=  \frac{1}{x_d(t)} \int_0^t\lambda(\bx(s))ds = \frac{(y(t) - y_k)}{x_d(t)}. 
\end{equation}
Using $\theta(\tau_k)=1$, we obtain  $x_d(t) = \lim_{t \to \tau_k^-}x_d(t) = \phi_k$. Thus, the fluctuations in $\phi_k$ and hence $z_k$ are ``hard-coded'' into the model via the kernel $k(\bx|\bx)$. This is a fundamental difference from the family of models considered above, where the dynamics of $z_k$ are emergent and must be obtained via some coarse-graining procedure that averages the growth dynamics over the cell-cycle. 

Note that by taking other hidden dimensions to be previous values of $\phi_k$ we may implement high order autoregressive dynamics for $\phi_k$.  In particular, we may take the kernel to have the general form
\begin{equation}
k(x_1,x_2,\dots,x_l|x_1',x_2',\dots,x_{l-1}') = \tilde{k}(x_1|x_1',x_2',\dots,x_{l-1}')\prod_{i=2}^{l-1} \delta(x_{i-1}-x_{i}')
\end{equation}
for some conditional density $\tilde{k}$.  
We will refer to this class of models as the fluctuating ratio (FR) model; however, we will not study the general FR model and will instead focus on a particular case that is more easily comparable to the $q$-SHO model.

\subsection{Coarse-grained correlations in minimal example}
We take an example with two noise sources, so the dynamics are
\begin{align}
\frac{d}{dt}y &= \lambda_0 + x\\
\frac{d}{dt}x_1 &= -ax_1 - f z + \sigma_1 \xi\\
\phi_k|y_k &= x_2|y_k  \sim {\rm Normal}(\phi_0 - \alpha y_k ,\sigma_2^2).
\end{align}
Here, the kernel is a gaussian and $x_1$ is a continuous scalar noise source analogous to $x_1$ in the $q$-SHO model. 
To be consistent with Eq. \ref{eq:frtheta}, we need
\begin{equation}
\frac{d}{dt}z = (\lambda_0 + x_1)\left(1 - \ln (2)/\phi_k \right). 
\end{equation}
The AR(1) process (Eq. \ref{eq:yar}) is an exact generative model for birth size dynamics in this setting. 

In \cite{levien2025size}, the case $f=0$ is studied and it is shown how this model can be written in a divisional formulation, where instead of specifying the distribution of $\phi_k$, the model is defined by a division rate $\beta(x,y,y_k)$ defined as 
\begin{equation}
\beta(x,y,y_k) = \lim_{dt \to 0}\frac{1}{dt}\bbP(\tau_k \in [t,t+dt)|\tau_k>t,y,y_k,x)
\end{equation}
where $y$ is the current size at time $t$. 
A key feature of the model when written in this manner is that the division rate factors as $\beta(x,y,y_k) = \lambda(x)\varphi(y,y_k)$; that is, the dependence on $x$ only appears through a multiplicative factor of $\lambda(x)$. This fact allows one to perform a distortion of the time axis that homogenizes the growth rates while preserving the coarse-grained size dynamics. Moreover, the size and growth are uncorrelated in the stationary distribution, similar to the position and velocity for the simple harmonic oscillator. 

The decoupling of size and growth implies $\alpha_{\lambda}=0$ when $f=0$. To see why, recall that $z_k$ may be centered without any loss of generality. Conditioning on the generation time and applying the tower property of expectation yields
\begin{align}
{\rm cov}(\bar{\lambda}_k,y_k) &= 
\E[\E[\bar{\lambda}_k,z_k|\tau_k]] = \E\left[\tau_k^{-1}\E\left[\int_0^{\tau_k}\lambda(s)dsy_k|\tau_k\right]\right]\\
&= \E\left[\tau_k^{-1}\int_0^{\tau_k}\E\left[\lambda(s)y_k|\tau_k\right]ds\right].
\end{align}
Note that after conditioning on $\tau_k$, $\lambda(s)$ and $y_k$ may only be correlated via correlations in the initial growth rate, but these correlations vanish when $f=0$.

   \begin{figure}[h!]
\centering
\includegraphics[width=0.5\textwidth]{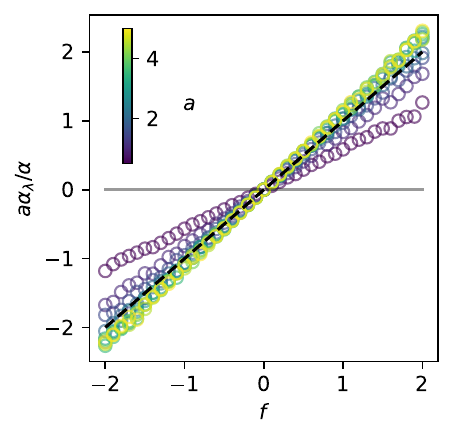}
\caption{The ratio $a\alpha_{\lambda}/\alpha$ for a range of $a$ values as a function of the feedback strength $f$. This should be compared to Figure \ref{fig:8}, where the ratio does not vanish when $q=0$. Other parameters are $\sigma_2^2=10^{-3}$ and $\alpha=1/2$.}\label{fig:9}
\end{figure}

Our claims are confirmed in Figure \ref{fig:9}, where we have shown a plot analogous to Figure \ref{fig:8} for this model. In this case, we use $f$, which controls the feedback from size to growth rates as the horizontal axis; in contrast to Figure \ref{fig:8}, the coarse-grained growth-size correlations vanish when there is no feedback.

\section{Discussion}
The question of how cell growth -- a measure of anabolic metabolism -- and the cell-cycle are coupled, as well as their role in maintaining size homeostasis, has become a central topic of interest in cellular physiology across all domains of life. In the field of cell-size regulation,  phenomenological models have proven to be useful tools for guiding experimental studies and establishing a framework for data analysis \cite{amir2014cell}.  In this paper, we have studied a continuous-time model with the goal of better understanding what coarse-grained correlations tell us about the coupling between the underlying processes; in particular, the structure of feedback from size to cell-cycle progression and growth. Our model can be viewed as an alternative to existing models, which attempt to treat cell-cycle progression and growth in a more symmetric manner.  It also has the advantage of being more explicitly connected to other models of homeostasis, since the discrete nature of cell-division can essentially be removed without influencing key statistical properties. 

We do not claim that the SHO type models are the ``correct'' description of real cell-size control; however, we find it helpful to illustrate the issues one encounters when interpreting coarse-grained correlations. This is illustrated by the discrepancy between the coarse-grained growth-size correlations and the mechanistic structure of feedback in the $q$-SHO model. This analysis, along with the comparison to the FR model, suggests some interesting avenues for future mathematical exploration in addition to the biological directions, such as a derivation of the coarse-grained correlations shown in Figure \ref{fig:8}.

From a biological perspective, it is interesting to compare the two models to experimental data. For such an analysis to be useful to biologists, more theoretical work is needed to understand what these models actually tell us about the underlying mechanisms by which metabolism and cell-cycle progression are regulated. A natural starting point is to consider a more biophysically motivated, yet still coarse-grained model, such as that in refs \cite{biswas2024single,willis2020limits}, and to derive the models considered here via model reduction procedures using phase reduction or spectral approximation.

We end by noting that the model introduced in this paper may be of interest beyond the size control community. Any homeostatic system where homeostasis is maintained  via the interplay of continuous and discrete corrections could potentially be related to this framework. In particular, we have in mind hormone circuits such as the HPA axis \cite{alon2023systems} which are subject to perturbations from a drug or surgery. 

\subsection*{Code availability}
All code to reproduce the figures in this paper can be found \href{https://github.com/elevien/GrowthCellCycleModel}{here}.

\subsection*{Acknowledgments}

Discussions with Peter Thomas, Naama Brenner, Kuheli Biswas and Ariel Amir greatly improved the quality of this manuscript. Levien was supported by the Burke Research Initiation at Dartmouth College and NSF DMS-2527337. 
 
\bibliographystyle{plain}
\bibliography{linear_model.bib}

@article{willis2020limits,
  title={Limits and constraints on mechanisms of cell-cycle regulation imposed by cell size-homeostasis measurements},
  author={Willis, Lisa and J{\"o}nsson, Henrik and Huang, Kerwyn Casey},
  journal={Cell Reports},
  volume={32},
  number={6},
  year={2020},
  publisher={Elsevier}
}

@book{williams2006gaussian,
  title={Gaussian processes for machine learning},
  author={Williams, Christopher KI and Rasmussen, Carl Edward},
  volume={2},
  number={3},
  year={2006},
  publisher={MIT press Cambridge, MA}
}

@article{tanouchi2017long,
  title={Long-term growth data of Escherichia coli at a single-cell level},
  author={Tanouchi, Yu and Pai, Anand and Park, Heungwon and Huang, Shuqiang and Buchler, Nicolas E and You, Lingchong},
  journal={Scientific data},
  volume={4},
  number={1},
  pages={1--5},
  year={2017},
  publisher={Nature Publishing Group}
}

@incollection{risken1989fokker,
  title={Fokker-planck equation},
  author={Risken, Hannes},
  booktitle={The Fokker-Planck equation: methods of solution and applications},
  pages={63--95},
  year={1989},
  publisher={Springer}
}

@article{witz2019initiation,
  title={Initiation of chromosome replication controls both division and replication cycles in E. coli through a double-adder mechanism},
  author={Witz, Guillaume and Van Nimwegen, Erik and Julou, Thomas},
  journal={Elife},
  volume={8},
  pages={e48063},
  year={2019},
  publisher={eLife Sciences Publications, Ltd}
}

@article{soifer2016single,
  title={Single-cell analysis of growth in budding yeast and bacteria reveals a common size regulation strategy},
  author={Soifer, Ilya and Robert, Lydia and Amir, Ariel},
  journal={Current Biology},
  volume={26},
  number={3},
  pages={356--361},
  year={2016},
  publisher={Elsevier}
}

@article{taheri2015cell,
  title={Cell-size control and homeostasis in bacteria},
  author={Taheri-Araghi, Sattar and Bradde, Serena and Sauls, John T and Hill, Norbert S and Levin, Petra Anne and Paulsson, Johan and Vergassola, Massimo and Jun, Suckjoon},
  journal={Current biology},
  volume={25},
  number={3},
  pages={385--391},
  year={2015},
  publisher={Elsevier}
}

@article{campos2014constant,
  title={A constant size extension drives bacterial cell size homeostasis},
  author={Campos, Manuel and Surovtsev, Ivan V and Kato, Setsu and Paintdakhi, Ahmad and Beltran, Bruno and Ebmeier, Sarah E and Jacobs-Wagner, Christine},
  journal={Cell},
  volume={159},
  number={6},
  pages={1433--1446},
  year={2014},
  publisher={Elsevier}
}

@article{luo2023stochastic,
  title={Stochastic threshold in cell size control},
  author={Luo, Liang and Bai, Yang and Fu, Xiongfei},
  journal={Physical Review Research},
  volume={5},
  number={1},
  pages={013173},
  year={2023},
  publisher={APS}
}

@article{knudson1971mutation,
  title={Mutation and cancer: statistical study of retinoblastoma},
  author={Knudson Jr, Alfred G},
  journal={Proceedings of the National Academy of Sciences},
  volume={68},
  number={4},
  pages={820--823},
  year={1971}
}

@article{lutkenhaus1993ftsz,
  title={FtsZ ring in bacterial cytokinesis},
  author={Lutkenhaus, Joe},
  journal={Molecular microbiology},
  volume={9},
  number={3},
  pages={403--409},
  year={1993},
  publisher={Wiley Online Library}
}

@book{alon2023systems,
  title={Systems medicine: physiological circuits and the dynamics of disease},
  author={Alon, Uri},
  year={2023},
  publisher={Chapman and Hall/CRC}
}

@article{biswas2024single,
  title={Single-cell growth rate variability in balanced exponential growth},
  author={Biswas, Kuheli and Sanderson, Amy E and Salman, Hanna and Brenner, Naama},
  journal={bioRxiv},
  pages={2024--06},
  year={2024},
  publisher={Cold Spring Harbor Laboratory}
}

@article{touchette2018introduction,
  title={Introduction to dynamical large deviations of Markov processes},
  author={Touchette, Hugo},
  journal={Physica A: Statistical Mechanics and its Applications},
  volume={504},
  pages={5--19},
  year={2018},
  publisher={Elsevier}
}

@article{karin2016dynamical,
  title={Dynamical compensation in physiological circuits},
  author={Karin, Omer and Swisa, Avital and Glaser, Benjamin and Dor, Yuval and Alon, Uri},
  journal={Molecular systems biology},
  volume={12},
  number={11},
  pages={886},
  year={2016}
}

@article{billman2020homeostasis,
  title={Homeostasis: the underappreciated and far too often ignored central organizing principle of physiology},
  author={Billman, George E},
  journal={Frontiers in physiology},
  volume={11},
  pages={200},
  year={2020},
  publisher={Frontiers Media SA}
}

@article{levien2025stochasticity,
  title={Stochasticity in mammalian cell growth rates drives cell-to-cell variability independently of cell size and divisions},
  author={Levien, Ethan and Kang, Joon Ho and Biswas, Kuheli and Manalis, Scott R and Amir, Ariel and Miettinen, Teemu P},
  journal={bioRxiv},
  pages={2025--06},
  year={2025},
  publisher={Cold Spring Harbor Laboratory}
}

@article{amir2014cell,
  title={Cell size regulation in bacteria},
  author={Amir, Ariel},
  journal={Physical review letters},
  volume={112},
  number={20},
  pages={208102},
  year={2014},
  publisher={APS}
}

@article{houzelstein2025generalized,
  title={Generalized dynamical phase reduction for stochastic oscillators},
  author={Houzelstein, Pierre and Thomas, Peter J and Lindner, Benjamin and Gutkin, Boris and P{\'e}rez-Cervera, Alberto},
  journal={Physical Review Research},
  volume={7},
  number={3},
  pages={033052},
  year={2025},
  publisher={APS}
}

@inproceedings{reith2023approximate,
  title={Approximate first-passage time distributions for Gaussian motion and transportation models},
  author={Reith-Braun, Marcel and Pfaff, Florian and Thummy, Jakob and Hanebeck, Uwe D},
  booktitle={2023 26th International Conference on Information Fusion (FUSION)},
  pages={1--8},
  year={2023},
  organization={IEEE}
}

@article{ho2018modeling,
  title={Modeling cell size regulation: From single-cell-level statistics to molecular mechanisms and population-level effects},
  author={Ho, Po-Yi and Lin, Jie and Amir, Ariel},
  journal={Annual review of biophysics},
  volume={47},
  number={1},
  pages={251--271},
  year={2018},
  publisher={Annual Reviews}
}

@book{amir2020thinking,
  title={Thinking Probabilistically: Stochastic Processes, Disordered Systems, and Their Applications},
  author={Amir, Ariel},
  year={2020},
  publisher={Cambridge University Press}
}

@article{kohram2021bacterial,
  title={Bacterial growth control mechanisms inferred from multivariate statistical analysis of single-cell measurements},
  author={Kohram, Maryam and Vashistha, Harsh and Leibler, Stanislas and Xue, BingKan and Salman, Hanna},
  journal={Current Biology},
  volume={31},
  number={5},
  pages={955--964},
  year={2021},
  publisher={Elsevier}
}

@article{anderson1990modeling,
  title={Modeling of solar oscillation power spectra},
  author={Anderson, Edwin R and Duvall Jr, Thomas L and Jefferies, Stuart M},
  journal={Astrophysical Journal, Part 1 (ISSN 0004-637X), vol. 364, Dec. 1, 1990, p. 699-705.},
  volume={364},
  pages={699--705},
  year={1990}
}

@article{hein2024asymptotic,
  title={Asymptotic decoupling of population growth rate and cell size distribution},
  author={Hein, Ya{\"\i}r and Jafarpour, Farshid},
  journal={Physical Review Research},
  volume={6},
  number={4},
  pages={043006},
  year={2024},
  publisher={APS}
}

@article{bjorklund2019cell,
  title={Cell size homeostasis: Metabolic control of growth and cell division},
  author={Bj{\"o}rklund, Mikael},
  journal={Biochimica et Biophysica Acta (BBA)-Molecular Cell Research},
  volume={1866},
  number={3},
  pages={409--417},
  year={2019},
  publisher={Elsevier}
}

@article{si2019mechanistic,
  title={Mechanistic origin of cell-size control and homeostasis in bacteria},
  author={Si, Fangwei and Le Treut, Guillaume and Sauls, John T and Vadia, Stephen and Levin, Petra Anne and Jun, Suckjoon},
  journal={Current Biology},
  volume={29},
  number={11},
  pages={1760--1770},
  year={2019},
  publisher={Elsevier}
}

@article{salman2025emergent,
  title={Emergent Homeostasis and Degeneracy from multi-Dimensional Attractors},
  author={Salman, Hanna and Biswas, Kuheli and Brenner, Naama},
  journal={bioRxiv},
  pages={2025--05},
  year={2025},
  publisher={Cold Spring Harbor Laboratory}
}

@article{levien2025size,
  title={Size-structured populations with growth fluctuations: Feynman--Kac formula and decoupling},
  author={Levien, Ethan and He{\"\i}n, Yair and Jafarpour, Farshid},
  journal={arXiv preprint arXiv:2508.14680},
  year={2025}
}

@article{wang2021regulation,
  title={Regulation of cell cycle progression by growth factor-induced cell signaling},
  author={Wang, Zhixiang},
  journal={Cells},
  volume={10},
  number={12},
  pages={3327},
  year={2021},
  publisher={MDPI}
}

@article{jones2001growth,
  title={Growth factor-dependent signaling and cell cycle progression},
  author={Jones, Steven M and Kazlauskas, Andrius},
  journal={Chemical Reviews},
  volume={101},
  number={8},
  pages={2413--2424},
  year={2001},
  publisher={ACS Publications}
}

@article{murray1991controls,
  title={What controls the cell cycle},
  author={Murray, Andrew W and Kirschner, Marc W},
  journal={Scientific American},
  volume={264},
  number={3},
  pages={56--65},
  year={1991},
  publisher={JSTOR}
}

@article{foreman2017fast,
  title={Fast and scalable Gaussian process modeling with applications to astronomical time series},
  author={Foreman-Mackey, Daniel and Agol, Eric and Ambikasaran, Sivaram and Angus, Ruth},
  journal={The Astronomical Journal},
  volume={154},
  number={6},
  pages={220},
  year={2017},
  publisher={IOP Publishing}
}

@article{10.1016/j.cub.2010.04.045,
    author = "Wang, Ping and Robert, Lydia and Pelletier, James F. and Dang, Wei and Taddéi, François and Wright, Andrew and Jun, Suckjoon",
    doi = "10.1016/j.cub.2010.04.045",
    title = "Robust Growth of Escherichia Coli",
    journal = "Current Biology",
    year = "2010"
}

@article{xia2020pde,
  title={PDE models of adder mechanisms in cellular proliferation},
  author={Xia, Mingtao and Greenman, Chris D and Chou, Tom},
  journal={SIAM journal on applied mathematics},
  volume={80},
  number={3},
  pages={1307--1335},
  year={2020},
  publisher={SIAM}
}

@article{10.7554/elife.88463,
    author = "Thiermann, Ryan and Sandler, Michael and Ahir, Gursharan and Sauls, John T. and Schroeder, Jeremy W. and Brown, Steven D. and Treut, Guillaume Le and Si, Fangwei and Li, Dongyang and Wang, Jue D. and Jun, Suckjoon",
    doi = "10.7554/elife.88463",
    title = "Tools and Methods for High-Throughput Single-Cell Imaging With the Mother Machine",
    journal = "Elife",
    year = "2024"
}

@article{cadart2018size,
  title={Size control in mammalian cells involves modulation of both growth rate and cell cycle duration},
  author={Cadart, Clotilde and Monnier, Sylvain and Grilli, Jacopo and S{\'a}ez, Pablo J and Srivastava, Nishit and Attia, Rafaele and Terriac, Emmanuel and Baum, Buzz and Cosentino-Lagomarsino, Marco and Piel, Matthieu},
  journal={Nature communications},
  volume={9},
  number={1},
  pages={3275},
  year={2018},
  publisher={Nature Publishing Group UK London}
}

@ARTICLE{10.1101/2023.11.23.568485,
  title     = "Scaling of stochastic growth and division dynamics: A
               comparative study of individual rod-shaped cells in the Mother
               Machine and {SChemostat} platforms",
  author    = "Ziegler, Karl F and Joshi, Kunaal and Wright, Charles S and Roy,
               Shaswata and Caruso, Will and Biswas, Rudro R and Iyer-Biswas,
               Srividya",
  journal   = "Mol. Biol. Cell",
  publisher = "American Society for Cell Biology (ASCB)",
  volume    =  35,
  number    =  6,
  pages     = "ar78",
  month     =  jun,
  year      =  2024,
  language  = "en"
}

@article{10.1371/journal.pone.0236534,
    author = "Seita, Akihisa and Nakaoka, Hidenori and Okura, Reiko and Wakamoto, Yuichi",
    doi = "10.1371/journal.pone.0236534",
    title = "Intrinsic Growth Heterogeneity of Mouse Leukemia Cells Underlies Differential Susceptibility to a Growth-Inhibiting Anticancer Drug",
    journal = "Plos One",
    year = "2021"
}

@article{lin2018homeostasis,
  title={Homeostasis of protein and mRNA concentrations in growing cells},
  author={Lin, Jie and Amir, Ariel},
  journal={Nature communications},
  volume={9},
  number={1},
  pages={4496},
  year={2018},
  publisher={Nature Publishing Group UK London}
}

\appendix

\section{Alternative models}\label{app:models}

\subsection{Relationship to inhibitor-dilution type models}

A common model for the triggering of cell division is the inhibitor dilution model. Here, cell-division is triggered when an inhibitor is diluted throughout the cell cycle drops below threshold. Let $c$ be the concentration of such an inhibitor, then $c_0e^{-\int \lambda ds}$. If $c_{\rm th}$ is the threshold then 
\begin{equation}
\int_0^{\tau} \lambda(s) ds = \ln c_0/c_{\rm th}
\end{equation}
A natural definition of $\theta$ is therefore
\begin{equation}
\theta =\frac{1}{ \ln c_0/c_{\rm th}} \int_0^{t} \lambda(s) ds = \frac{1}{ \ln c_0/c_{\rm th}}(y-y_0)
\end{equation}
since division occurs when $\theta=1$. 
Hence, in the most basic version of this model with $\ln c_0/c_{\rm th}$ is held constant, $y$ and $\theta$ are directly coupled. 

To connect to our models, we take the time derivative of $\theta$: 
\begin{equation}
\frac{d}{dt}\theta =  \nu =  \frac{1}{ \ln c_0/c_{\rm th}}\lambda. 
\end{equation}
If we allow $c_0$ and $c_{\rm th}$ to vary, this model fits naturally into the framework of the fluctuating ratio model of Section \ref{sec:noage}. However, we can introduce an independent source of noise which decouples $\theta$ and $y$. This comes from defining 
\begin{equation}
\frac{d}{dt}\theta =  \nu(\bx) =\alpha(\bx)\lambda(\bx). 
\end{equation}
Our linear model emerges when we expand to leading order in $\bx$: 
\begin{equation}
\frac{d}{dt}\theta =  \alpha(0)\lambda(0) + \alpha(0)\nabla \lambda(0) ^T\bx
+  \lambda(0) \nabla \alpha(0)^T\bx + \cdots 
\end{equation}
Therefore, the interpretation of $\bnu_{x}$ would be 
\begin{equation}
\bnu_{x} = \nabla (\alpha(\bx)\lambda(\bx))\Big|_{\bx =0} = \alpha(0)\nabla \lambda(0) ^T + \lambda(0) \nabla \alpha(0)^T 
\end{equation}

\section{Eigenvalue problem in the general case}\label{app:evalgeneral}
In general, when $A$ has different eigenvalues, the stability of the deterministic dynamics are determined by the eigenvalues of
\begin{equation}
\widetilde{M}_z = \begin{bmatrix} 
0 &  \brho_{x}^T\\
-\bff_{y} & -A 
\end{bmatrix}.
\end{equation}
Applying the Shur formula 
\begin{equation}
\det(\widetilde{M}_z -\mu I)= \det(-A-\mu I)\left(-\mu + \brho_x^T(A - \mu I)^{-1}\bff_y \right)
\end{equation}
If $\mu$ is an eigenvalue of $-A$ the first term is zero, but then $A + \mu I$ is not invertible. Only when the corresponding eigenvector is orthogonal to $\blambda_x$ does  this eigenvalues correspond to eigenvalues of $\tilde{M}_z$. The other eigenvalues are solutions to \begin{equation}\label{eq:mueq}
-\mu + \brho_x^T(A - \mu I)^{-1}\bff_y= 0. 
\end{equation}
From the perspective of stability analysis we may always assume that $A = \Lambda = {\rm diag}(\ba)$ is a diagonal matrix, as any model of this form can be recast as such with a redefinition of $\brho_x$ and $\bnu_{\theta}$.\footnote{When addressing questions about the stationary distribution we need to remember that this the diagonalization also transforms the noise matrix; however, in the small noise limit this will not influence whether size homeostasis is maintained as long as the noise matrix is not degenerate.} 
Therefore, the characteristic polynomial of $\widetilde{M}$ (Eq. \ref{eq:mueq}) can always be rewritten in the form 
\begin{equation}\label{eq:mu12_general}
- \mu - \sum_{i=1}^d \frac{\rho_{x,i}f_{y,i}}{a_i + \mu} = 0, 
\end{equation}
We can say something about the solution when $a_i = a + \epsilon_i$, for $\langle \epsilon_i \rangle = 0$ and $v_{\epsilon} = \langle \epsilon_i^2 \rangle \ll 1$ (here $\langle \cdot \rangle$ is the average over dimensions.) The equation polynomial can be approximated as
\begin{equation}
-\mu (a + \mu)^3 - (a + \mu)^2 \omega_0^2 + \omega_0^2 v_{\epsilon} =0
\end{equation}
Because the derivative is positive at the root when $v_{\epsilon} =0$, the term $\omega_0^2 v_{\epsilon}$ will always increase the magnitude of the dominant negative eigenvalue. Small variation in the relaxation rate between hidden dimensions will increase the rate of convergence. 

\end{document}